# Web Usage mining framework for Data Cleaning and IP address Identification

Priyanka Verma
The IIS University
Jaipur, Rajasthan

Dr.Nishtha Kesswani
Central University of Rajasthan
Bandar Sindri,Ajmer

*Abstract—* *The World Wide Web is the most wide known information source that is easily available and searchable. It consists of billions of interconnected documents Web pages are authored by millions of people. Accesses made by various users to pages are recorded inside web logs. These log files exist in various formats. Because of increase in usage of web, size of web log files is increasing at a much faster rate. Web mining is application of data mining technique to these log files. It can be of three types Web usage mining, Web structure mining and Web content mining. Web Usage mining is mining of usage patterns of users which can then be used to personalize web sites and create attractive web sites. It consists of three main phases: Preprocessing, Pattern discovery and Pattern analysis. In this paper we focus on Data cleaning and IP Address identification stages of preprocessing. Methodology has been proposed for both the stages. At the end conclusion is made about number of users left after IP address identification.*

## I. INTRODUCTION

Web site is a group of web pages. Web pages may contain text, images, and videos. They are linked by hyperlinks through which navigation happens. Whenever user accesses any website, log files are created. Log file records entire information about each user's website access. Due to increase in usage of web sites the size of log files is increasing day by day. Data stored in Web log files can exist in various formats such as the NCSAs Common log file format, the W3C Extended Log File format or the IIS log file format. There are different kinds of log files which includes Error logs, Referrer logs, and Access logs. Log files are created in various locations like web server, proxy server, and Client browser. For getting optimum results we need to extract data from all three log files. Analysis of the patterns of user's habits and interests helps in increasing performance of web site, by improving web site design. Web mining is the application of various data mining techniques to discover data from web documents and services. Web mining is divided into three types Web content mining, Web structure mining and Web usage mining. Web Content Mining deals with the discovery of information from the contents or data or documents or services of web. Web Structure Mining mines the structure of hyperlinks within the website. Web Usage Mining mines the usage data stored in the logs. Web usage mining analyzes information about web pages which were navigated by users. Analysis of such information helps us to discover the unknown and potentially interesting patterns. Web usage mining is also called as Click-stream analysis.

## II. WEB USAGE MINING

Web usage mining (WUM) or web log mining, is revealing user's behavior or usage patterns by applying various data mining techniques on data stored in web log file. The ability to know the patterns of user's habits and interests helps in efficient building of various web based applications. The main source of data for web usage mining consists of various logs stored on numerous web servers, web clients all around the world. A clickstream is the aggregate sequence of page visits executed by a particular user navigating through a Web site. In addition to page views, clickstream data also consist of cookies, metatags, and other data used to transfer web pages from server to browser. Web Usage mining is mining of this clickstream data stored in logs.

### A. Stages of Web Usage Mining

There are four stages of web usage mining

1. Data Collection: During this phase data is collected from various log files stored in various locations.

2. Preprocessing: This is the most important phase, it takes 80% of the effort of entire web usage mining process. During this phase data is extracted from logs files collected is cleaned and then users are identified and then sessions are made from users identified.

3. Pattern Discovery: This phase discovers various patterns followed by the user.

4. Pattern analysis: After patterns are discovered from web logs analysis is done using various query mechanism such as SQL to perform various OLAP operations

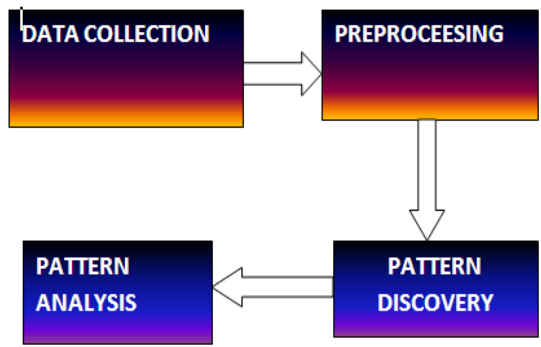

Fig. 1. Stages of Web Usage Mining

III. PREPROCESSING STAGE OF WEB USAGE MINING

Whenever loading a particular web page, the browser also requests for all the objects embedded in the page, such as .gif or .jpg graphics files. These requests for gif and jpeg files are also logged in log files. These requests have to be cleaned before any further analysis of user patterns. Thus click stream data stored in log files requires substantial preprocessing before user behavior can be analyzed. Preprocessing is second phase after data collection, and most essential step before discovery and analysis. Input to preprocessing phase is log file. It is large in size and contains number of raw and irrelevant entries and therefore cannot be directly used in Web usage mining process. Preprocessing of log fie is a very complex and painstaking job and it takes 80% of the total time of web usage mining process as whole.

A. Log File

Log file is a plain text file which records information about each user which includes name, IP address, date, time, and bytes transferred, access request. Web server writes information in log file each time a user requests a resource from that particular site. When user submits request to a web server that activity is recorded in web log file. Log file size ranges from 1KB to 100MB. Log files give us information about:

1. Pages requested in website

2. Bytes sent from server to user

3. Type of error occurred

B. Location of Log Files

Log files are located in three different locations given below.

1.Server : These log files record usage of data of web server. These log file do not record cached pages visited. Data of these log files contains sensitive and personal information so web server keeps them closed.

2.Proxy server: Web proxy server is intermediary between user and web server, it takes HTTP requests from user, gives them to web server, then result is passed from web server and returned to user. Client send request to web server via proxy server.

3.Client browser: These log files reside in client's browser window itself. This information can be recorded only if cookies are enabled. Cookies are pieces of information generated by a web server and stored in user's computer

C. Log File formats

Data in log files exists in three different formats
- W3C Extended log file format

This format is default log file format on IIS server. Fields are separated by space. Time is recorded as GMT (Greenwich Mean Time). Year is recorded as YYYY-MM-DD. It contains fields:
- Software used
- Version of Software
- Date and Time of access
- IP Address of user
- Method URI stream
- Status of protocol
- Version of protocol

```
#Software: Microsoft Internet Information Services 7.5
#Version: 1.0
#Date: 2012-02-05   06:57:20
#Fields: date time cs-method cs-uri-stem c-ip cs-version sc-status
    2012-01-09   3:56:27 GET /WebSite/ ::1 HTTP/1.1 301
```

Fig. 2. W3C Extended log file [2]

- NCSA(National Center For Supercomputing Application) common log file format

This format records basic information about user's request such as user name and remote host name, date, time, request type, HTTP status code and numbers of bytes send by server. It is fixed format and cannot be customized. Year is in format DD/MMM/YYYY. Fields are separated by space.

```
::1 - - [19/Jan/2012:10:00:30 +0530] "GET /Website/ HTTP/1.1" 200 1107
```

Fig. 3. NCSA log file [2]

- IIS log file format

This format again cannot be customized, it is fixed format. Fields are separated by comma. Time is recorded in local time. Fields are
- Client IP address

- User name
- Date and time of access
- Service and instance
- Server name and server IP address
- Time taken for request to be responded
- Bytes sent by client
- Bytes sent by server
- Service status code and windows status code
- Type of request
- Target of operation and parameters

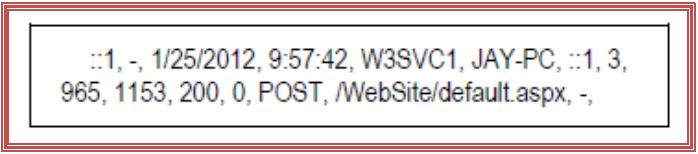

Fig. 4. IIS log file [2]

*D. Stages of Preprocessing*

In order to get the suitable Web log data to perform the mining to discover useful patterns, we must undertake a series of operations (Preprocessing) on the original Web log files such as data cleaning, user and transaction identification, data integration and so on. [5].The first stage of preprocessing is Data cleaning is also known as data cleanup. Data cleaning is a process in which data errors, inconsistencies are removed. The data cleaning process removes the noise or irrelevant data which includes jpeg, gif files or script images. The size of log file, after successful completion of the data cleaning step is reduced [6].Next step of preprocessing is User Identification, For user identification firstly, if the IP addresses are identical, but the browser software or the operating systems are different in Agent information, then two different users can be assumed. Secondly, if IP address and the Agent information are identical, then we should judge whether there is a connection between the pages which are requested to access and the pages which have been accessed. If there is no direct link between them, then we can assume that there exist multiple users in the machine which accesses the Web sites [5]. Next stage is Session identification. Session is a sequence of requests made by a single user with a unique IP address on a Web site during a specified period of time. There are two approaches proposed by [7] and [8] for session identification. First approach is Smart-SRA (Smart Session construction) Algorithm which is designed to overcome deficiencies of the time and the navigation oriented heuristics. Smart-SRA uses a novel approach to construct user session as a set of paths in the web graph where each path corresponds to users' navigations among web pages [7]. In other approach given by GuerBas et al.[8] the original time oriented sessionistic algorithm is modified to make sure that identified sessions from the same sequence share a pattern i.e. there should be some specified pages in common between sessions [8]. Next stage is Path Completion which is a process of adding the page accesses that are not in the web log but those which have actually occurred. These Missing page references occur due to caching [9] and can be completed through path completion

## IV. EXISTING WORK

A lot of work on preprocessing stage of web usage mining is ongoing. Algorithms for certain stages of web usage mining including Data Cleaning, User Identification, and Session Identification have been proposed but they have a few problems which have to be addressed.

## V. PROPOSED PREPROCESSING METHODOLOGY

The main motivation behind performing web usage mining is that the information stored in World Wide Web is increasing day by day and so is the user's demand to get right data. For getting right data user's navigational patterns have to be studied. For this various mining techniques have to be used which will help us to extract meaningful patterns and relationship from large data [10].In the field of web usage mining researchers have done considerable research but rapid development of internet makes these studies lag behind. In this paper we have performed preprocessing using log file which initially contains 500 access records. Software has been created to perform Data Cleaning and User Identification using algorithms proposed by [2] for data cleaning, Field extraction and [11] for user identification using VB.NET and data extracted is stored in SQL server.

*A. Field Extraction*

During this stage fields are extracted from log file containing 500 records and is stored in excel file. The separator used for field extraction is space character. Records extracted in MS Excel are then stored in SQL server to perform data cleaning and user identification. After field extraction extracted records as stored in SQL table log table are shown below. Fields extracted from excel file are URL name and IP address. It also includes SNO field to find out number of records extracted.

*B. Data cleaning*

Data cleaning is second stage after the storage of log file in table. This stage is performed to remove the unnecessary content which includes requests for images, styles and scripts or other files. This stage is the most simplest of all stages as it consists of only filtration of the data. During this stage all URLS with jpeg, gif and .css extensions are removed using algorithm proposed by [2].After data cleaning number of records is reduced to 441 from 500 records. There is not much reduction in size of file as 500 records that were taken contained very few URLs with .jpeg, gif and .css extensions

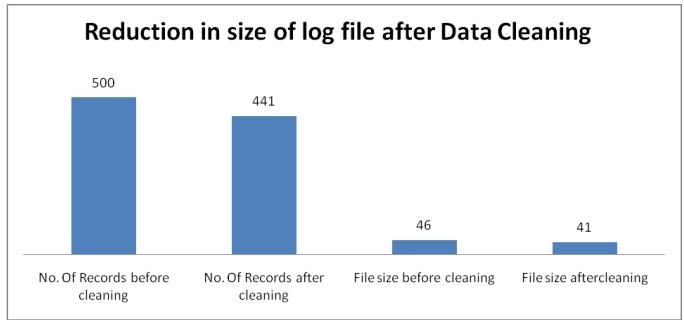

### C. User/IP Address identification

During this stage unique users having same IP address are identified using algorithm given by [11].Logic used is "If IP address is same but browser version or operating system is different then it represents different user." [9].After IP address identification number of users identified is 52.This means out of 500 records extracted from log file containing various URLs accessed by the users, there are 52 unique users.

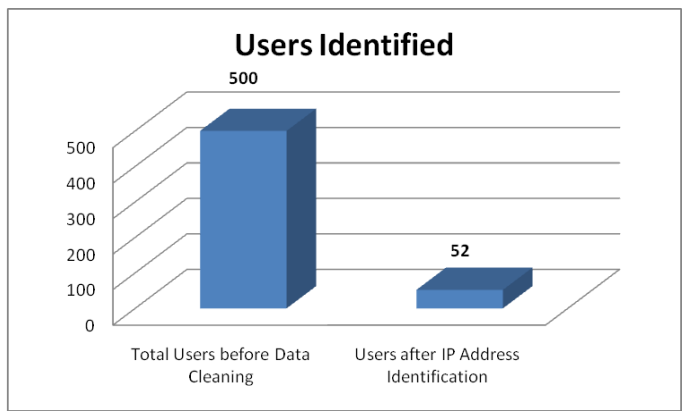

### CONCLUSION

Log files contain data which has to be cleaned before any user identification. Preprocessing of log files is one of the most important steps of web usage mining. In this paper we have performed two stages of preprocessing Data Cleaning and User Identification using algorithms proposed by [2] and [11].We have observed that after performing data cleaning on log file containing 500 records number of records, size of files reduces though not drastically as URLs taken do not contain much unnecessary content such as .jpeg,.css extension files. Further after user identification stage of preprocessing number of users identified is 52 out of 441 records obtained after Data Cleaning. This shows that out of 441 user records of URL's accessed there are 52 unique users.

### FUTURE WORK

Performing session identification and then Path Completion. Once all the preprocessing stages are completed certain patterns would be revealed which when analyzed will help us to conclude on access patterns followed by user and this data will help us to personalize web sites according to the user.